\documentclass[conference]{IEEEtran}
\IEEEoverridecommandlockouts
\usepackage{cite}
\usepackage{amsmath,amssymb,amsfonts}
\usepackage{nicefrac}
\usepackage{algorithmic}
\usepackage{algorithm2e}
\usepackage{graphicx}
\usepackage{textcomp}
\usepackage{xcolor}
\usepackage{comment}
\usepackage[]{todonotes}
\usepackage{tikz}
\usepackage{hyperref}
\usepackage{cleveref}
\usepackage{subcaption}
\usepackage{booktabs}
\usepackage{tabularx}
\usepackage[autolanguage]{numprint}
\usetikzlibrary{shapes}

\definecolor{sender}{HTML}{88c0d0}
\definecolor{receiver}{HTML}{81a1c1}
\newtheorem{definition}{Definition}
\newtheorem{lemma}{Lemma}
\newtheorem{theorem}{Theorem}

\newtheorem{remark}{Remark}

\def\BibTeX{{\rm B\kern-.05em{\sc i\kern-.025em b}\kern-.08em
    T\kern-.1667em\lower.7ex\hbox{E}\kern-.125emX}}
    
\newcolumntype{Y}{>{\centering\arraybackslash}X}
\begin{document}

\title{2PPS -- Publish/Subscribe with Provable Privacy
\thanks{This work was supported by funding of the German Research Foundation (DFG), research grant 317688284 and by funding of the Helmholtz Association (HGF) through the Competence Center for Applied Security Technology (KASTEL)}}

\makeatletter
\newcommand{\linebreakand}{%
  \end{@IEEEauthorhalign}
  \hfill\mbox{}\par
  \mbox{}\hfill\begin{@IEEEauthorhalign}
}
\makeatother

\author{
    \IEEEauthorblockN{Sarah Abdelwahab Gaballah}
    \IEEEauthorblockA{TU Darmstadt\\\texttt{gaballah@tk.tu-darmstadt.de}}
    \and
    \IEEEauthorblockN{Christoph Coijanovic}
    \IEEEauthorblockA{Karlsruhe Institute of Technology\\\texttt{christoph.coijanovic@kit.edu}}
    \and
    \IEEEauthorblockN{Thorsten Strufe}
    \IEEEauthorblockA{Karlsruhe Institute of Technology\\\texttt{thorsten.strufe@kit.edu}}
    \linebreakand
    \IEEEauthorblockN{Max M\"uhlh\"auser}
    \IEEEauthorblockA{TU Darmstadt\\\texttt{max@tk.tu-darmstadt.de}}
}

\maketitle

\begin{abstract}
Publish/Subscribe systems like Twitter and Reddit let users communicate with many recipients without requiring prior personal connections. The content that participants of these systems publish and subscribe to is typically public, but they may nevertheless wish to remain anonymous. While many existing systems allow users to omit explicit identifiers, they do not address the obvious privacy risks of being associated with content that may contain a wide range of sensitive information.

We present 2PPS (\emph{Twice-Private Publish-Subscribe}), the first pub/sub protocol to deliver strong provable privacy protection for both publishers and subscribers, leveraging Distributed Point Function-based secret sharing for publishing and Private Information Retrieval for subscribing.
2PPS does not require trust in other clients and its privacy guarantees hold as long as even a single honest server participant remains.
Furthermore, it is scalable and delivers latency suitable for microblogging applications.

A prototype implementation of 2PPS can handle \numprint{100000} concurrent active clients with 5 seconds end-to-end latency and significantly lower bandwidth requirements than comparable systems.

%Our evaluation of  results show that 2PPS can support \numprint{100000} active clients with less than $5$ seconds end-to-end latency which is faster by a factor of $> 400$ than Riposte and Blinder. 
%Also, the end-to-end communication costs in 2PPS are $300\times$ less than Pung, Riposte, and Blinder.

%Publish/Subscribe systems let users communicate with many recipients without requiring prior personal connections.
%Metadata, such as which topics a given client is subscribed to often reveals sensitive information (e.g., health data) to service-providers and adversaries.
%We present 2PPS (\emph{Twice-Private Publish-Subscribe}), the first pub/sub system that offers strong provable privacy protection for both publishers and subscribers while supporting hundreds of thousands of clients with latency suitable for microblogging applications.
%2PPS requires no trust in other clients and only a single honest server to provide its privacy guarantees.
%It is based on Distributed Point Function-based secret sharing for publishing and Private Information Retrieval for subscribing. 
%Our evaluation results show that 2PPS can support 100,000 active clients with less than $5$ seconds end-to-end latency which is faster by a factor of $> 400$ than Riposte and Blinder. 
%Also, the end-to-end communication costs in 2PPS are $300\times$ less than Pung, Riposte, and Blinder.
\end{abstract}


\section{Introduction} \label{sec:in}

Consider the immense popularity of services like Twitter, Reddit, and Telegram.
All can be classified as implementing the Publish/Subscribe (pub/sub) messaging pattern:
Messages are \emph{published} to certain \emph{topics} (e.g., hashtags for Twitter or channels for Telegram).
Users can freely \emph{subscribe} to topics they are interested in and will receive corresponding messages. 
The service acts as an intermediary broker between publisher and subscribers and is responsible for managing subscriptions, sorting received messages by topic, and forwarding them to the intended subscribers.

One particularly interesting use of pub/sub systems is the organization of volunteering, political involvement, and activism.
Pub/sub lends itself to this setting since it allows large numbers of people to connect without having a prior personal relationship.
Protesters in Iraq, Hong Kong, and Belarus have been using Telegram and FireChat for this purpose~\cite{RefFC1, RefFC2, RefTG1}.
However, the use of conventional systems can leak valuable \emph{metadata} to an adversary:
\begin{itemize}
    \item An activist who is found out to be publishing or subscribing to a regime-critical topic may be facing serious legal consequences.
    \item If the regime finds out how many users are subscribed to a critical topic, it can determine the size of the activists' movement and deploy an overwhelming police force at the next protest.
\end{itemize}
Telegram and FireChat do not protect this kind of metadata~\cite{RefFC3}.
This is where our proposed protocol, 2PPS, comes in:
It offers strong provable privacy protection for both publishers and subscribers in an open pub/sub setting.
History shows that a state-level adversary can have access to immense resources~\cite{RefNSABudget}.
Thus, we aim to protect against an adversary who may not only corrupt users and servers but also observe and interfere with traffic globally.
While the example of political activists impressively motivates the need for private pub/sub communication, it is not the only possible use for 2PPS.
Service providers can infer sensitive information such as health problems, financial status, or sexual preferences by observing which topics a user is active in.

At a high level, 2PPS reaches its goal as follows:
The broker’s functionality is distributed over multiple servers, where privacy is ensured as long as at least one arbitrary server does not collude with the adversary.
While the adversary inherently learns which messages are published, since she can join arbitrary groups herself, she cannot learn any further information (e.g., who publishes which message).
This is achieved by using \emph{Distributed Point Function} (DPF)-based secret sharing.
Compared to prior work~\cite{RefRiposte, RefExpress}, we present an improved secret sharing approach that also protects against active interference. 
Subscribers protect their privacy by using \emph{Private Information Retrieval} (PIR) for receiving messages.

To the best of our knowledge, no existing protocol can provide open pub/sub communication with strong provable privacy guarantees for both publishers and subscribers. 
Some protocols require trusted group members \cite{RefDissent, RefTalek, RefM2} or trusted execution environments~\cite{RefPubsubsgx}.
Others don't provide both sender and receiver anonymity \cite{RefM2} or don't target worst-case protection~\cite{Refgiakkoupis}.
Additionally, some of them are vulnerable to traffic analysis attacks \cite{RefMTor}.
PIR-based protocols offer strong cryptographic security guarantees and hide metadata efficiently \cite{RefExpress, RefRiposte, RefBlinder}.
However, the majority of these protocols support either point-to-point communication \cite{RefExpress} or broadcasting \cite{RefRiposte, RefBlinder}.
PIR-based protocols that support selective multicast communication either provide strong receiver anonymity but weak sender anonymity \cite{RefTalek}, or do not scale well \cite{RefRiffle}.

Designing an anonymous communication protocol always requires a trade-off between privacy protection, trust, and overhead~\cite{RefBoundsSoK}.
However, we show that 2PPS, despite strong privacy protection and minimal trust requirements, manages to keep the overhead for clients at a reasonable level:
\begin{itemize}
\item It incurs a latency of \numprint[s]{25} to handle one million users where each client submits a \numprint[B]{160} message and receives a \numprint[KB]{10} block of messages.
\item For \numprint{50} subscriptions per client, 2PPS requires $\numprint{300}\times$ less bandwidth compared to broadcasting systems such as Riposte \cite{RefRiposte} and Blinder \cite{RefBlinder}.
\end{itemize}

\paragraph*{Contributions} In this paper, we make the following contributions:
\begin{itemize}
\item We introduce 2PPS, a new anonymous publish/subscribe protocol by combining private writing using Distributed Point Functions (DPF) and private reading using Information-Theoretic Private Information Retrieval (IT-PIR).
\item We give formal proof to show that the 2PPS protocol reaches our stated goals of Publisher- and Subscriber Unobservability.
\item We provide an evaluation of 2PPS that demonstrates its efficiency in terms of latency and bandwidth.
\end{itemize}
\section{Model and Goals}
\label{sec:goal}
\subsection{Protocol Overview}
2PPS implements the publish/subscribe model:
When \emph{publishing} a message, the sending user (i.e., the publisher) specifies a \emph{topic} the message belongs to.
The protocol then delivers this message to all \emph{subscribers} of this topic. 
We assume an open environment, meaning users can freely subscribe to and unsubscribe from any topic they wish.

The 2PPS network consists of $n$ clients and $N$ servers with $n \gg N$. Each server stores a full copy of two databases, a \emph{write} database $D_w$ and a \emph{read} database $D_r$ .  All servers collectively maintain the contents of these databases.
The read database $D_r$ is partitioned into a set of $\ell_r$ equal-sized blocks, with a publicly known topic for each block.
The write database $D_w$ is partitioned into $\ell_w$ equal-sized blocks, each sized to hold one message.

Similar to related protocols~\cite{RefRiffle, RefRiposte, RefDissent2}, communication in 2PPS occurs in rounds to defend against traffic analysis attacks. Each round is split into three distinct phases (\Cref{fig:ov} depicts these phases at a high level). During the first phase, each client deposits exactly one message into the write database $D_w$, which is shared among the $N$ servers using secret sharing.
If a client has no real message to send, it generates a cover message.
After the first phase has concluded, the servers collaborate to reveal all messages simultaneously.
During the second phase, cover messages are discarded and the remaining messages are sorted into their corresponding topic-block of the read database.
Finally, in the third phase, clients anonymously retrieve the messages from their subscribed topic.

\begin{figure}
    \centering
    \begin{tikzpicture}[font=\scriptsize,scale=0.9]
        \node[draw,circle,fill=sender] (c0) at (-2.5,1.5) {\texttt{A}};
        \node[draw,circle,fill=sender] (c1) at (-2.5,0.5) {\texttt{B}};
        \node[draw,circle,fill=sender] (c2) at (-2.5,-0.5) {\texttt{C}};
        \node[draw,circle,fill=sender] (c3) at (-2.5,-1.5) {\texttt{D}};
        \node[draw,circle,fill=receiver] (c4) at (5.25,1) {\texttt{E}};
        \node[draw,circle,fill=receiver] (c5) at (5.25,0) {\texttt{F}};
        \node[draw,circle,fill=receiver] (c6) at (5.25,-1) {\texttt{G}};

        \node[anchor=west] (m0) at (-1,1) {$t_1:m_1$};
        \node[anchor=west] (m1) at (-1,0) {$t_1:m_2$};
        \node[anchor=west] (m2) at (-1,-0.5) {$t_3:m_3$};
        \node[anchor=west] (m3) at (-1,-1.5) {$t_2:m_4$};

        \draw[] (-1,1.75) rectangle (0.2,-1.75);
        \draw[] (-1,1.25) --++(1.2,0);
        \draw[] (-1,0.75) --++(1.2,0);
        \draw[] (-1,0.25) --++(1.2,0);
        \draw[] (-1,-0.25) --++(1.2,0);
        \draw[] (-1,-0.75) --++(1.2,0);
        \draw[] (-1,-1.25) --++(1.2,0);
        \node[] (l0) at (-0.4,-2) {$D_w$};

        \node[anchor=west] (t0) at (2,1) {$t_0:-$};
        \node[anchor=west] (t1) at (2,0.5) {$t_1:m_1,m_2$};
        \node[anchor=west] (t2) at (2,0) {$t_2:m_4$};
        \node[anchor=west] (t3) at (2,-0.5) {$t_3:m_3$};
        \node[anchor=west] (t4) at (2,-1) {$t_4:-$};

        \draw[] (2,1.25) rectangle (4,-1.25);
        \draw[] (2,0.75) --++(2,0);
        \draw[] (2,0.25) --++(2,0);
        \draw[] (2,-0.25) --++(2,0);
        \draw[] (2,-0.75) --++(2,0);
        \node[] (l1) at (3,-1.5) {$D_r$};

        \draw[->] (0.2,1) to[out=0,in=180] (2,0.5);
        \draw[->] (0.2,0) to[out=0,in=180] (2,0.5);
        \draw[->] (0.2,-0.5) to[out=0,in=180] (2,-0.5);
        \draw[->] (0.2,-1.5) to[out=0,in=180] (2,0);

        \draw[dashed,->] (c0) to[out=0,in=180] (-1,-0.5);
        \draw[dashed,->] (c1) to[out=0,in=180] (-1,-1.5);
        \draw[dashed,->] (c2) to[out=0,in=180] (-1,1);
        \draw[dashed,->] (c3) to[out=0,in=180] (-1,0);

        \draw[dashed,->] (4,-1) to[out=0,in=180] (c4);
        \draw[dashed,->] (4,0.5) to[out=0,in=180] (c5);
        \draw[dashed,->] (4,0) to[out=0,in=180] (c6);
        
        \draw[dashed,black!50] (-0.4,3.6)--(-0.4,1.75);% (l0.center)--(-0.4,-2.27);
        \draw[dashed,black!50] (3,3.6)--(3,1.25); %(l1.south)--(3,-2.26);

        \node[anchor=north,align=center] (l2) at (-1.75,3.6) {
            \textbf{Phase 1}\\
            Private publishing\\
            of messages with\\
            included topic id\\
            to random slots
        };
        \node[anchor=north,align=center] (l3) at (1.3,3.6) {
            \textbf{Phase 2}\\
            Revealing messages\\
            and grouping them by\\
            topics into $D_r$
        };
        \node[anchor=north,align=center] (l4) at (4.7,3.6) {
            \textbf{Phase 3}\\
            Private retrieval of\\
            messages of subscribed\\ 
            topics
        };
    \end{tikzpicture}
    \caption{
        Server databases ($D_w$ and $D_r$) and general protocol flow.
    }
    \label{fig:ov}
\end{figure}
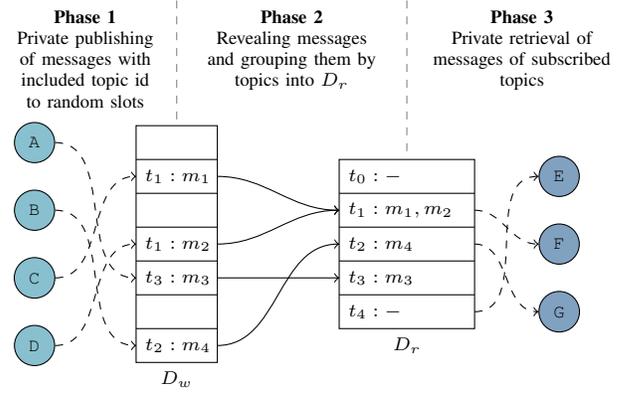

\subsection{Threat Model}
2PPS assumes a strong adversary $\mathcal{A}$, whose goal is to compromise the privacy of honest users.
$\mathcal{A}$ is assumed to be in control of all network links.
Thus, she may not only passively observe all traffic on every link, but also insert, delay, drop, time, and modify arbitrary packets.
Further, $\mathcal{A}$ may corrupt $N-1$ servers and an arbitrary number of clients.
Since we assume open groups, $\mathcal{A}$ may join all groups as a subscriber through corrupted clients.
We assume that honest clients and servers behave as specified by the 2PPS protocol.

\subsection{Security Goals}
\label{sec:goals:notions}
2PPS aims to achieve the formal privacy notions of \emph{Publisher Unobservability} and \emph{Subscriber Unobservability}.
Kuhn et al. present a set of game-based privacy notions for unicast communication~\cite{RefNotions}, which we adapt for the pub/sub scenario.

Each notion is defined by a game played between a challenger $\mathcal{C}$ and an adversary $\mathcal{A}$:
\begin{enumerate}
    \item $\mathcal{C}$ chooses a random challenge bit $b\in\{0,1\}$.
    \item $\mathcal{A}$ submits a challenge consisting of two self-chosen scenarios $(S_0,S_1)$ to $\mathcal{C}$.
        Each secenario contains a number of communications $(p,m,t)$, where $p$ denotes the publisher, $m$ the message, and $t$ the topic that $p$ sends $m$ to.
    \item $\mathcal{C}$ checks the received challenge for validity and, if valid, simulates the protocol execution of the communications contained in $S_b$.
    \item Based on his abilities, $\mathcal{A}$ gets to observe and interact with the protocol execution.
    \item $\mathcal{A}$ determines which of his scenarios was chosen and submits his guess $b^\prime\in\{0,1\}$ to $\mathcal{C}$.
        $\mathcal{A}$ wins if $b=b^\prime$.
\end{enumerate}
Steps 2-5 can be repeated.
Instead of a challenge, the adversary can also submit a \emph{subscription update}.
The subscription update specifies for each client the topics she is subscribed to in each scenario.
If differences in the communications between the two batches lead to differences in protocol behavior that $\mathcal{A}$ can observe, then $\mathcal{A}$ gains an advantage over randomly guessing the chosen batch.

A concrete privacy notion defines which information is allowed to leak to the adversary and which should be protected by the protocol.
Information that is allowed to leak may not differ between batches to ensure that $\mathcal{A}$ does not gain an unfair advantage when trying to distinguish.
If $\mathcal{A}$ can still determine which of his batches was submitted to the protocol with a non-negligible advantage over random guessing, the protocol has failed to protect the information it was supposed to protect and therefore does not reach this privacy notion.

\paragraph*{Publisher Unobservability}
With publisher unobservability, the protocol aims to hide any information about active publishers.
Implicitly, information about messages, topics, and subscribers is not protected.
Thus, $\mathcal{A}$ is required to submit batches that only differ in the publisher of each communication:
Le the $i$th communication of the submitted first batch be $(\mathbf{p_0^i},t_0^i,m_0^i)$ where topic $t_0^i$ has subscribers $r_0^{i,0},\dots,r_0^{i,n}$.
Then, the $i$th communication of the submitted second batch has be of form $(\mathbf{p_1^i},t_0^i,m_0^i)$ where $t_0^i$ also has to have subscribers $r_0^{i,0},\dots,r_0^{i,n}$.
If $\mathcal{A}$ can determine which of his batches was executed with a non-negligible advantage despite this restriction, the protocol does not reach publisher unobservability. 
\paragraph*{Subscriber Unobservability}
Analogous to the publisher variant, subscriber unobservability aims to hide any information about active subscribers.
Since information about publishers, messages, and topics is allowed to leak, $\mathcal{A}$ has to submit the \emph{same} communications in both batches.
However, which clients are subscribed to which topics may vary between the two batches of a challenge.
\section{2PPS Architecture}
\label{sec:arch}
This section describes the 2PPS protocol in greater detail.
\Cref{sec:arch:p1} presents the anonymous publishing phase, \Cref{sec:arch:p2} the management of published requests and \Cref{sec:arch:p3} the anonymous subscription phase.

\subsection{Phase 1: Anonymous Publishing.}
\label{sec:arch:p1}
Assume that client Alice wants to publish some message $m$ to topic $t$ without the adversary being able to link the message to her.
We start by introducing a simple but inefficient method to hide sender identities from malicious servers.
Then we improve the efficiency of this method and finally extend it to also protect against adversaries that are in control of the network.

Na\"ively, \emph{secret sharing}~\cite{RefSSS} can be employed for anonymous publishing:
First, Alice computes a vector $w$ of the same length as the write database $D_w$, which contains $(t\mid m)$ at a random index and 0 everywhere else.
She then computes $N$ secret shares $w_1,\dots,w_N$ with the following properties:
\begin{enumerate}
    \item $\sum_{i=1}^N w_i = w$
    \item Any combination of $N-1$ secret shares does not reveal any information about $(t\mid m)$ or the index at which $(t\mid m)$ is located.
\end{enumerate}
Alice distributes the shares to the servers, where the $i$th server $S_i$ receives $w_i$.
$S_i$ then adds $w_i$ to its read database state $D_{w}^i$:
\begin{align*}
    D_w^i \gets D_w^i + w_i
\end{align*}

To hide sending frequencies, all clients are required to publish \emph{exactly} one message per round.
If the client has no ``real'' message to send, she may send a message consisting only of zeroes to a random topic as cover.
After processing requests from multiple clients, the servers can collaborate to compute a combined database $D_w = \sum_{i} D_w^i$.
As long as every client chose a unique index for her messages, $D$ contains all original messages.

This approach is quite inefficient:
For every write request, a vector with the same size as the database has to be sent.
To address this issue,  Riposte~\cite{RefRiposte} suggested the use of distributed point functions (DPF).
\begin{definition}[DPF]
    Let $f_{i^\ast,m} : \{0,\dots,\ell_w\} \mapsto \mathbb{F}$ be a point function with
    \begin{align*}
        f_{i^\ast,m}(i) &= \begin{cases}
            m &\text{for } i = i^\ast\\
            0 &\text{for } i\in\{0,\dots,\ell\}\setminus i^\ast
        \end{cases}
    \end{align*}
    $f_A, f_B : \{0,\dots,\ell_w\} \mapsto \mathbb{F}$ are distributed point functions of $f_{i^\ast,m}$, iff
    \begin{enumerate}
        \item Neither $f_A$ nor $f_B$ by themselves reveal anything about $m$ or $i^\ast$
        \item $\forall\: i\in \{0,\dots,\ell\} : f_A(i) + f_B(i) = f_{i^\ast,m}(i)$
    \end{enumerate}
\end{definition}

DPFs can be used to \emph{compress} the shares sent to the servers from the na\"ive approach:
Alice runs $\textsf{GenDPF}(t\mid m)$, which generates $N$ DPF-shares $f_1,\dots,f_N$ that contain $(t\mid m)$ at a random index.
These shares are distributed among the servers, with server $S_i$ receiving $f_i$.
Server $S_i$ can derive $w_i$ by evaluating $f_i(j)$ at every point $j\in\{0,\dots,\ell_w\}$.
Current research states that sending a DPF share instead of $w_i$ directly reduces the communication cost to $O(\lambda\cdot\log \ell_w + \log |(t\mid m)|)$ bits where $\lambda$ is the security parameter~\cite{RefFSS}.

As is, this approach protects against malicious servers, but not against stronger adversaries, who are also in control of the network:
$\mathcal{A}$ could simply intercept Alice's shares before they reach the servers and combine them to reveal Alice's topic and message.
To protect, Alice can encrypt each share with the receiving server's public key before sending it.
Since we assume at least one honest server, $\mathcal{A}$ cannot gain access to all shares.
Due to the public nature of messages in 2PPS, there are further active attacks possible:
\paragraph*{Replay}
To link Alice to her message by replay, $\mathcal{A}$ saves all shares Alice sends in a given round and all messages that are revealed in the same round.
In the next round, $\mathcal{A}$ inserts the saved shares into the traffic.
the servers will add them to their $D_w$ state.
$\mathcal{A}$ can identify which messages Alice has sent by observing which identical message was revealed in both rounds.
To detect replayed messages, Alice includes a current timestamp with every share (inside of the encryption layer).
An honest server can check the time stamp for currentness and refuse further participation in this round if this check fails.
The share-encryption also prevents $\mathcal{A}$ from selectively modifying the timestamp.
$\mathcal{A}$ could also replay the shares in the same round as the original shares, which would corrupt Alice's message.
However, this attack is easily detectable by the honest server, since it has access to all shares of the current round at once and can check for duplicates.
\paragraph*{Modification}
Our proposed protection against replay attacks also enables honest servers to detect if a received request was modified by $\mathcal{A}$.
We assume that encryption used to protect the share and the timestamp provides \emph{diffusion}, i.e., ensures that any change of a ciphertext leads to widespread and unforeseeable changes of the plaintext.
Thus, any modification of a request leads to a significant change of the included timestamp with overwhelming probability.
The honest server detects this invalid timestamp and refuses further participation in the current round.
\paragraph*{Drop \& Delay}
Attacks based on dropping and delaying messages are both very common and hard to avoid in anonymous communication~\cite{RefDrop}.
If a powerful adversary can drop the requests of all but one client, then he can unambiguously link this client to her messages once the shares are combined.
A less powerful adversary might have insider knowledge of a message that will be sent in a given round.
If he drops the request of the suspected sender and the message is not published, his suspicion is confirmed.

To detect a dropped request, the honest server needs to know how many messages are supposed to arrive in a given round.
Related literature commonly assumes that protocol participation is static, i.e., that clients are always online \cite{RefVuvuzela, RefPung}.
This is a very strong and arguably not very realistic assumption since it discards user churn.
We introduce an additional mechanism that enables us to make weaker assumptions regarding client participation:

The \emph{verifiable participation commitment} requires every client who joins the network to send a message to each server with which the client commits herself to participate in the next $k$ rounds.
The parameter $k$ can be chosen by the client to fit his routine.
A client could for example join when arriving at his office in the morning and commit to participating until his usual end of the workday.
Clients can also commit to shorter periods and renew their commitment periodically.
With that, the honest server knows from which clients to expect requests in any given round.
If fewer requests than expected are received, the server assumes that a malicious drop must have occurred and refuses further participation to protect the senders' privacy.
The adversary also needs to be prevented from replacing dropped requests with self-generated ones to circumvent the protection.
This can be done by requiring the clients to include a \emph{digital signature} with every request.
That way, each request can be linked to the client who sent it and the server can \emph{verify} that all committed clients have indeed participated.

\subsection{Phase 2: Managing Published Messages.}
\label{sec:arch:p2}
When the writing epoch ends, the servers reveal the published messages among each other by combining their $D_w$ states.
As a first step, all cover messages (i.e., those which only contain zeroes) are discarded.
Together, the servers choose a block size for $D_r$ such that every topic's messages fit into a single block.
Thus, the block size may be changing from round to round depending on the number of messages per topic, but at any point, all blocks of $D_r$ have the same size.
Next, the servers append each message in $D_w$ to its corresponding topic-block in $D_r$.
These messages are stored temporarily in $D_r$ until the end of the communication round.

\paragraph*{Updating the Topic List}
Over time, new topics will be created and others will become inactive, thus there is a need for periodically updating the list of current topics and their corresponding blocks in $D_r$.
To create a new topic, the client follows the same steps as when sending a message to an existing topic but includes a new topic id.
The servers take notice of the unknown id and save the included message.
During the next database update, a block for this new topic is created, and the topic is included in the list of topics sent to the clients.
The first message from the original creator is added to this topic in the following round. 
Regarding deleting the inactive topics from $D_r$, servers will consider a topic as inactive, if there are no published messages on this topic for some configured number of rounds.
The topic list and mapping from topic to block ID are updated periodically and clients are informed of the update afterward.

\subsection{Phase 3: Anonymous Subscribing.}
\label{sec:arch:p3}
2PPS allows clients to anonymously subscribe to topics and get new messages without polling them. 
Like related protocols~\cite{RefRiffle, RefTalek}, it depends on information-theoretic PIR (IT-PIR).
The anonymous subscription consists of two building blocks:
Subscription registration and message retrieval.

\paragraph*{Private Subscription Registration}
2PPS requires all clients to update their subscriptions at a fixed rate to hide changes in interest.
Every time a client receives a topic-list update, he has to renew his subscription.
If the client is not interested in any topic, he sends a subscription request to a random topic.
Clients that newly join the network need to wait until the next topic update to start participation.

Assume that Alice wants to subscribe to the $j$th topic.
To do so, she creates a vector $q\in \{0,1\}^{\ell_r}$, which is equal to 1 at position $j$ and equal to 0 at all other positions.
Alice then computes a subscription request $req_i = Enc_{pk_i}(s_i|q_i)$ for each server $S_i$ with $i\in\{1,\dots,N\}$.
$Enk_{pk_i}(\cdot)$ is an encryption under the $S_i$'s public key $pk_i$, $s_i\in \{0,1\}^{\ell_w}$ is a randomly chosen shared secret and the PIR query $q_i$ is computed as follows:
\begin{align*}
    q_i &= \begin{cases}
        \text{random } & \text{for } i < N\\
        q \oplus q_1 \oplus \dots \oplus q_{K-1} & \text{for } i = N
    \end{cases}
\end{align*}
The shared secret $s_i$ is locally updated each round synchronously at client and server (e.g., using a cryptographic
hash function or a key schedule).

To reduce the client's inbound bandwidth, related literature~\cite{RefTalek, RefRiffle} suggests the use of a random server $P$ as a proxy for the client:
Instead of sending the subscription requests $q_1,\dots,q_N$ directly to the servers, the client sends them to his proxy $P$.
$P$ forwards these requests to corresponding servers where they are stored. 

\begin{remark}[Multiple Subscriptions]
    Each subscription registration may only contain a subscription to a single topic.
    If a client wants to be subscribed to multiple topics simultaneously, he has to send multiple subscription requests.
    To hide the number of topics clients are subscribed to, all clients need to send the same number of subscription requests.
    These can contain a mix of real subscriptions and cover subscriptions to random topics.
\end{remark}

\paragraph*{Private Messages Retrieval}
In every round, each server $S_i$ computes a response $res_i$ for each stored subscription by taking the XOR of all $D_r$ blocks that have 1 in their positions in the PIR query.
Instead of sending the responses directly to the client, the servers submit them to $P$ who computes $res \gets \bigoplus_{i\in\{1,\dots, N\}} res_i$, and forwards it to the client.
Thus, the client's incoming bandwidth is reduced by a factor of $N$.
To prevent $P$ from learning which topic the client has subscribed to, each server has to obfuscate its response.
Server $S_i$ obfuscates its response by computing $res_i \gets res_i \oplus s_i$, the client can restore the desired block of published messages by computing $res \oplus s_1 \oplus \dots \oplus s_N$.

\section{Analysis of 2PPS security properties}
\label{sec:ana}
In this section, we show that 2PPS reaches our formalized privacy goals as defined in \Cref{sec:goals:notions}.
\begin{theorem}[Publisher Unobservability]
    \label{thm:pub}
    2PPS achieves Publisher Unobservability.
\end{theorem}
Intuitively, reaching Publisher Unobservability requires unlinking senders from their messages and hiding which senders are active.
To unlink senders from their messages, 2PPS employs secret sharing based on distributed point functions.
Each server receives a secret share that does not reveal any information about the contained message by itself.
Only once all clients have submitted their shares, the combination of all shares is revealed all at once.
We strengthen the secret sharing scheme against adversaries in control of the whole network by introducing a timestamp and an additional layer of encryption around the shares.
This prevents the adversary from being able to modify or replay shares.
Further, we introduce the verifiable participation commitment which enables honest servers to detect dropped shares.
Finally, we hide which senders are active by requiring all clients to send at a fixed rate, creating cover messages when they don't have a real message to send.
A full proof of security can be found in \Cref{sec:app:so}.

\begin{theorem}[Subscriber Unobservability]
    \label{thm:sub}
    2PPS achieves Subscriber Unobservability.
\end{theorem}
In 2PPS, IT-PIR ensures that subscribers cannot be linked to the topics they're subscribed to.
Both subscription requests and the responses containing the messages appear random to any adversary that is not in control of either the client itself or \emph{all} servers.
The use of cover traffic and synchronized round further ensures that the adversary cannot gain any information about the frequencies at which the client updates his subscription or receives messages.
A full proof of security can be found in \Cref{sec:app:ro}.

\section{Performance Evaluation}
\label{sec:eval}
Privacy protection should not be only restricted to those with access to powerful hardware, but also it should support users who have limited bandwidth and computational power, e.g., in a mobile setting.
2PPS aims to keep the overhead at a reasonable level to enable as many clients as possible to participate.
The goal of our evaluation is to investigate the impact of using DPFs (for private writing), and IT-PIR (for private reading) together on computation and network overhead on both client and server sides. 
One important measure of scalability is the end-to-end latency of a system.
For 2PPS, we evaluate the influence of a changing number of participating clients, the number of subscribed topics per client, and the number of messages per topic on the latency of the system.

\paragraph*{Implementation}
A prototype of our protocol is implemented in C and Go. 
We use Go for the high-level operations of client and server.
Cryptographic primitives are used from the DEDIS advanced crypto library and Go's native crypto library.
We rely on the available source code of Express\footnote{\url{https://github.com/SabaEskandarian/Express}} for C implementations of the auditing protocol and DPFs.
To update the shared secrets between clients and servers locally, we use keyed AES similar to Riffle~\cite{RefRiffle}. 
We conduct the experiments on three virtual machines, each equipped with a 16-core Intel Xeon E5-2640 v2 processor and \numprint[GB]{64} of RAM. 
All three machines are located in the same data center. 
We operate two of them as servers and use the third one to simulate the clients. 
In all experiments, each client is configured to send one message per round (LS in \Cref{fig:eva} refers to the length of the sent message). 
All clients participate in every round. 
Also, we test the performance of private retrieval for three different block sizes: 10 KB, 64 KB, and 256 KB (LR in \Cref{fig:eva} refers to the length of the retrieved block). We adopt these block sizes from \cite{RefPung}, \cite{RefExpress} and \cite{RefRiffle}.  

\paragraph*{Baselines} 
We compare 2PPS to three different PIR-based anonymous group communication protocols: Riposte \cite{RefRiposte}, Pung \cite{RefPung}, and Blinder \cite{RefBlinder}. 
We choose these protocols since they provide cryptographic anonymity guarantees similar to 2PPS. Riposte and Blinder support anonymous broadcasting, whereas Pung and 2PPS provide selective multicast communication (i.e., they allow the users to fetch only the messages that are interesting to them).

\subsection{Computation Overhead}
To understand the computation costs that are imposed on the client and server-side, we run a set of experiments in which every client sends one \numprint[KB]{1} message and retrieves one \numprint[KB]{64} block per round. 
Between experiments, we vary the number of messages processed by the servers (i.e., the number of participating clients). 
Since each client selects a random row to write its message into, collisions are possible and lead to the irreversible corruption of both colliding messages.  In this experiment, we use a large fixed database $D_w$ to handle this issue. This database achieves write success rates of \numprint[\%]{99.8}, \numprint[\%]{98}, and \numprint[\%]{82} for \numprint[]{e3}, \numprint[]{e4}, and \numprint[]{e5} messages respectively (according to the success rate formula in \cite{RefRiposte}). As the size of $D_w$ is fixed, clients and servers pay fixed CPU costs for each write-request regardless of the number of messages received by the servers. 

As shown in \Cref{tab:table1}, the client’s operations are all comparatively inexpensive.
Note that, ``Create PIR query'' is done only during subscription updates rather than every round.
``Process PIR reply'' denotes the time it takes to XOR the received PIR reply with the shared secrets to reveal the messages. 
The computation costs for the servers are dominated by the first phase (``Expand DPF shares'' and the ``Audit''). 
Also, the second phase introduces overhead which can be broken down further into the costs of two sub-phases: combining the shared $D_w$ states, which accounts for the largest part of the overall time of the second phase, and the grouping of messages into their corresponding topics.

\begin{table}
\begin{tabularx}{\columnwidth}{lYYY}\toprule
 & \multicolumn{3}{c}{\textbf{\# Messages Processed}} \\
 & \numprint[]{e3} & \numprint[]{e4} & \numprint[]{e5} \\
 \midrule
\multicolumn{4}{l}{\textbf{Client CPU costs}}\\
Generate DPF shares & \numprint[\mu s]{229.26} & \numprint[\mu s]{229.26} & \numprint[\mu s]{229.26}\\
Audit & \numprint[\mu s]{204.16} & \numprint[\mu s]{204.16} & \numprint[\mu s]{204.16}\\
Create PIR query & \numprint[\mu s]{0.533} & \numprint[\mu s]{6.73} & \numprint[\mu s]{12.54}\\
Process PIR reply & \numprint[\mu s]{19.33} & \numprint[\mu s]{21.79} & \numprint[\mu s]{22.027}\\
 \addlinespace
\multicolumn{4}{l}{\textbf{Server CPU costs}}\\
Expand DPF shares & \numprint[s]{5.62} & \numprint[s]{5.62} & \numprint[s]{5.62}\\
Audit & \numprint[ms]{36.52} & \numprint[ms]{36.52} & \numprint[ms]{36.52} \\
Second Phase & & & \\
\textit{1. Combine $D_w$ states} & \numprint[ms]{13,27} & \numprint[ms]{14,01} & \numprint[ms]{19.66}\\
\textit{2. Group messages} & \numprint[ms]{0.08} & \numprint[ms]{0.87} & \numprint[ms]{6.70}\\
Process PIR Query & \numprint[ms]{0.17} & \numprint[ms]{1.19} & \numprint[ms]{10.06}\\
\bottomrule
\end{tabularx}
\caption{
    Cost of 2PPS operations under varying the number of messages stored on the server where the size of each message is 1 KB. 
    } \label{tab:table1}
\end{table}

\subsection{Network Overhead}
In this section, we discuss the network overhead of 2PPS. 
These measures are especially important in bandwidth-restricted scenarios, such as mobile communication.
We split our discussion into two parts:
First, we consider the total bandwidth consumed by the operation of 2PPS versus related protocols.
Second, we investigate how much ``unnecessary'' bandwidth in form of cover messages is required for 2PPS. 

\paragraph*{Comparative bandwidth comsumption}
\Cref{fig:ntotalcost} shows the total communication cost to send one message and retrieve one block by the client when the number of participating clients varies. 
That includes all the sent and received messages between the client and the server to achieve private writing and reading. 
Since we use a fixed size for $D_w$, the communication costs of private writing (including the auditing) are not growing when the number of clients increases.

Compared to Pung, 2PPS's anonymous writing is more expensive.
However, Pung's reading phase introduces a high communication overhead that makes the total cost of communication using Pung significantly more expensive than 2PPS.  
For instance, when there are one million messages on the server, Pung has a total communication cost that is $\numprint{1263}\times$ larger than 2PPS to send a \numprint[KB]{1} message and retrieve a \numprint[KB]{10} block. 
Note that, in the shown results, the costs of Pung include sending the PIR query by the client to retrieve the messages. 
If we assume that the PIR queries are stored already on the servers (similar to 2PPS), this will reduce the Pung costs to $\numprint{950}\times$ larger than 2PPS which is still a substantial difference.
Pung's high overhead occurs because clients do not know the index of the data that they are interested in. 
Instead, they perform a binary search on the database through multiple CPIR queries. 
Further, the size of the CPIR answer increases as the CPIR recursion depth becomes higher. 
Angel et al. state that Pung's approach results in an increase of network overhead by a factor of $\log(n)$ for a database of $n$ elements compared to PIR with known indices~\cite[Section 7.4]{RefPung}. 

Riposte also requires communication costs much higher than 2PPS. 
For \numprint[]{e6} clients, 2PPS has $\numprint{5033}\times$ less total cost than Riposte.  
2PPS has better performance for two reasons: 1) it uses a new generation of DPFs \cite{RefDPF} and an auditing protocol \cite{RefExpress} that is more efficient than the one used in Riposte; 2) it sends to each client only a subset of the published messages, whereas Riposte broadcasts all messages to all clients. 

Blinder introduces high total communication costs due to its need to operate by a large number of servers to achieve its anonymity guarantees (in our experiments, we ran Blinder on 5 servers). 
Additionally, it broadcasts all published messages to every client resulting in high communication overhead similar to Riposte. 
For \numprint[]{e6} clients, the total network cost for each Blinder client is around \numprint[MB]{160} while the corresponding value in 2PPS is about \numprint[KB]{32}. 

To further explore the performance implications of using IT-PIR on the network download overhead, \Cref{fig:nnetdown1} depicts the total amount of data a client receives during 20 rounds for one subscription per client. 
In 2PPS, the amount of data per round equals the number of subscriptions a client has times the block size. 
A client who subscribes to one topic with a block size of \numprint[KB]{64}, downloads \numprint[MB]{1.28} during 20 rounds. 
Pung introduces a much higher communication overhead than 2PPS as well, but less than broadcasting. 
A Pung user downloads more data than a 2PPS user by a factor of $\numprint{54}-\numprint{512}\times$ to retrieve a \numprint[KB]{10} block.

To compare the download overhead in 2PPS with a broadcast-based approach such as Riposte and Blinder, we consider broadcasting under three different assumptions regarding the number of cover messages. 
The first case (denoted as ''broadcasting \numprint[\%]{100}'') assumes that all messages that clients sent to the servers are real messages. 
Analogously, broadcasting \numprint[\%]{50} and \numprint[\%]{25} assume that \numprint[\%]{50} and \numprint[\%]{25} of messages are real, respectively. 
Note that cover messages are discarded and therefore not broadcast to the clients. 
For one million clients, the network download of broadcasting (\numprint[\%]{100}) is \numprint[GB]{20} in 20 rounds, resulting in a $\numprint{15625}\times$ increase over 2PPS when the block size is \numprint[KB]{64}.  
Hence, this method is extremely inefficient in terms of bandwidth for popular services.

As shown in \Cref{fig:nnetdown2}, 2PPS also considerably outperforms the three broadcasting variants when the number of subscriptions per client increases. 
Therefore, adopting IT-PIR in our protocol to retrieve the interesting messages generally allows bandwidth-efficient communication between client and server. 
That makes 2PPS suitable for bandwidth-restricted users.

\begin{figure*}[t!]
    \centering
            \begin{subfigure}[t]{0.32\textwidth}
        \includegraphics[height=48mm]{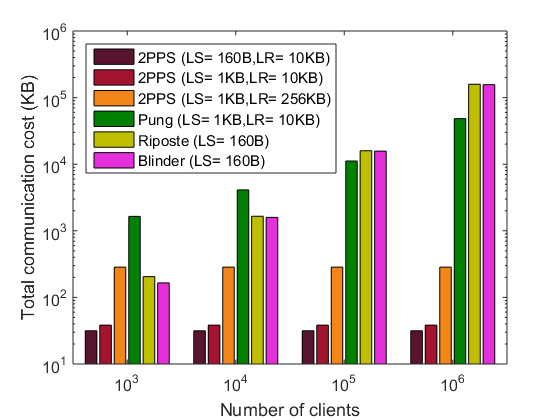}
        \caption{Total communication costs when the client sends one message and retrieves one block of messages.}\label{fig:ntotalcost}
    \end{subfigure}
    \begin{subfigure}[t]{0.32\textwidth}
        \includegraphics[height=48mm]{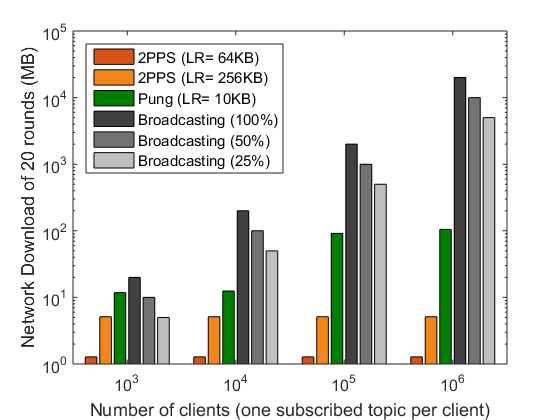}
        \caption{The total amount of downloaded data by a client after 20 rounds with varying numbers of clients and block sizes.}\label{fig:nnetdown1}
    \end{subfigure}
    \begin{subfigure}[t]{0.32\textwidth}
        \centering
        \includegraphics[height=48mm]{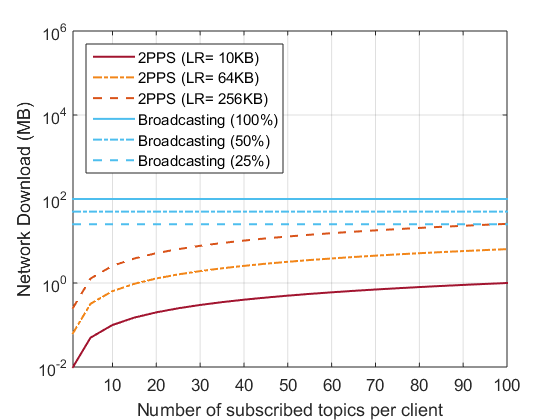}
        \caption{The total amount of downloaded data by a client for \numprint{100000} clients with varying numbers of subscribed topics and block sizes.}\label{fig:nnetdown2}
    \end{subfigure}
           \begin{subfigure}[t]{0.32\textwidth}
        \centering
        \includegraphics[height=48mm]{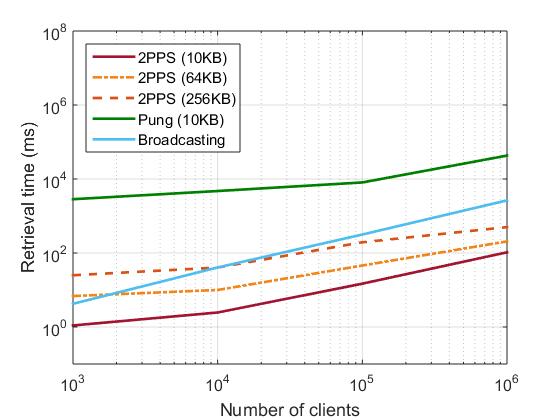}
        \caption{Retrieval time of one block per client with varying number of clients and block size.}\label{fig:ntime}
    \end{subfigure}
        \begin{subfigure}[t]{0.32\textwidth}
        \centering
        \includegraphics[height=48mm]{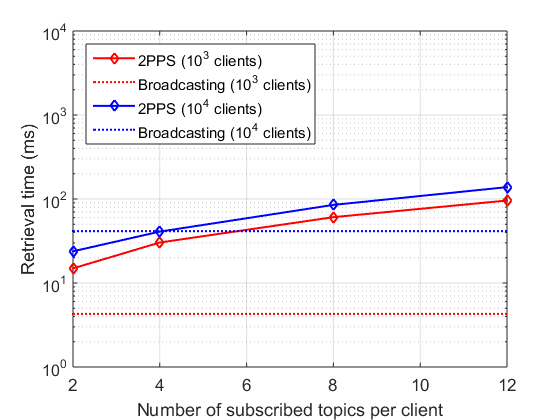}
        \caption{Retrieval time when a client subscribes to different number of topics.}\label{fig:ntime2}
    \end{subfigure}
    \begin{subfigure}[t]{0.32\textwidth}
        \centering
        \includegraphics[height=45mm]{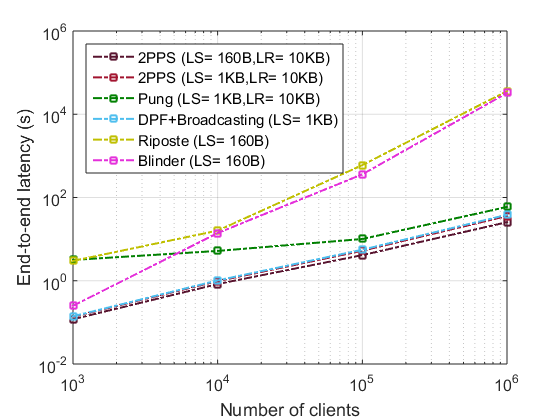}
        \caption{End-to-end latency of message delivery when a client subscribes to one topic.}\label{fig:ntime1}
    \end{subfigure}

    \caption{Evaluation Results}
    \label{fig:eva}
\end{figure*}

\paragraph*{Average Cover Ratio}
While 2PPS's retrieval method introduces no overhead for downloading a fixed amount of data, another kind of overhead occurs if topics have differing popularity:
Similar to the ``Counting Bound'' proposed by Gelernter and Herzberg~\cite{RefLimit}, hiding which topic a client is subscribed to \emph{requires} a certain amount of overhead:

Assume that $\mathcal{A}$ knows the combined size of all messages sent to each topic and assume that each client subscribes to a single topic.
To hide any information about the link between clients and their subscribed topics, \emph{all} clients have to receive a response large enough that it could contain all messages from the most popular topic.
Otherwise, $\mathcal{A}$ can eliminate topics from the list of possible subscriptions for a given client which have more messages than the client receives.

To fulfill the Receiver Counting Bound, 2PPS pads each topic to the same size as the most popular topic using cover massages;
a client subscribed to any topic that receives fewer messages than the most popular one will receive some amount of cover messages with his PIR response creating network overhead.
We want to evaluate how much of a client's request on average consists of cover messages.
While we have no real usage data for 2PPS, we can find related literature to approximate how topic popularity might be distributed:
According to Chen et al.~\cite{RefTPop}, the popularity distribution of Twitter hashtags follows Zipf's law with an exponent of $\alpha \approx \numprint{0.8}$.
Further, Liu et al.~\cite{RefRSS} have analyzed the RSS feed characteristics and determined that the popularity of feeds sorted by the number of requests also follows Zipf's law with an exponent of $\alpha\approx \numprint{1.37}$.
We use Chen et al.'s result to approximate the number of messages sent to a given topic and Liu et al.'s result to determine the likelihood of a client subscribing to a given topic.
In our experiment, we assumed that the most popular topic receives 1000 messages and that the $i$th most popular topic receives $\nicefrac{1}{i^{0.8}}$ as many messages.
We chose a random topic index $j$ according to a Zipfian distribution with parameter $\alpha = \numprint{1.37}$ and determined how many cover messages topic $j$ required.
Under these assumptions, clients receive approximately \numprint[\%]{54} cover messages per retrieval on average.
Note that this number highly depends on the actual usage patterns of the service:
If all topics receive the same number of messages, \emph{no} cover messages are required, if topic popularity varies widely, more than \numprint[\%]{54} cover messages might be received on average.

\subsection{End-to-End Latency}  % update the figures of this subsection
To evaluate the efficiency of our protocol, we are interested in computing the total time required to send one message and retrieve one block by each client. 
The latency in 2PPS is measured as the total time of all three protocol phases. 
The~time of DPF evaluation represents the expensive part of the total latency, and it depends on the size of $D_w$. 
Having a fixed large database means a fixed big evaluation time for DPF shares even when the number of received write requests is small.  
For less latency, the evaluation time can be reduced by changing the size of $D_w$ based on the number of expected requests. However, the database should be still large enough to handle requests successfully with high probability.

The latency of the third phase is determined by the number of topics, the block size, and the number of subscriptions per client. 
The more topics the read database contains or the larger each topic block is, the longer it takes to compute a PIR reply. 
If a client subscribes to more topics, the number of PIR replies that need to be computed increases, increasing latency. 

\Cref{fig:ntime} illustrates how the retrieval time of one block scales whith varying  block sizes and numbers of clients. 
In general, retrieving messages using either our method or broadcasting doesn’t cause much latency. 
We compared our protocol to broadcasting to understand the performance implication of using PIR for distributing messages to the receivers instead of using a broadcast. 
2PPS is considerably faster than Pung, even when retrieving larger blocks.

\Cref{fig:ntime2} illustrates the increase in retrieval time when a 2PPS client subscribes to many topics. 
The number of participating clients influences the time required to broadcast messages more than it does with 2PPS. 
That is especially true when most of the published messages are real. For \numprint{e4} clients, retrieving 12 blocks using 2PPS takes \numprint[ms]{97} more than the broadcasting, which means the difference in the latency time between the two approaches is comparatively insignificant. 
Therefore, 2PPS's retrieval method can reduce the bandwidth without leading in return to high latency, even if the clients have many subscriptions.

\Cref{fig:ntime1} shows the end-to-end latency for posting one message when we vary the number of clients. 
Again, this figure demonstrates the negligible effect of retrieval time on the total latency, as the end-to-end latency of 2PPS is slightly smaller than the method that relies on DPF and broadcasting (instead of IT-PIR). 
2PPS has significantly better performance than Riposte as it adopts a more efficient DPF version~\cite{RefDPF} and auditing method~\cite{RefExpress}. 

We compared the latency of 2PPS to Pung when sending one \numprint[KB]{1} message and retrieving one \numprint[KB]{10} block (10 messages in the block). 
As shown in \Cref{fig:ntime1}, 2PPS outperforms Pung especially for a modest number of clients. 
In Pung, communication partners are required to trust each other, which enables clients to agree upon secret mailboxes for their message exchange. 
Since the link between clients and their mailboxes is not known to the adversary, messages can be directly written to the mailboxes, resulting in much less overhead than our DPF-based approach. 
The use of a single untrusted server in Pung requires much more overhead in the retrieval of messages than IT-PIR based approach of 2PPS. 
Overall, this expensive reading phase more than offsets any advantages in the latency of Pung's writing phase. 
The latency of Blinder is close to Riposte and considerably higher than 2PPS. Similar to 2PPS, the most expensive part in Blinder is the process of expanding the submitted blind write requests to add them to the database. For \numprint{e5} clients, the total time to serve this number of clients is around 8 minutes, and the expanding process acquires \numprint[\%]{96} of the total latency. 2PPS supports the same number of clients with less than half of Blinder's time.

%\cc{We should also compare 2PPS against Aloha (\url{https://eprint.iacr.org/2021/044.pdf}), which is similar to Pung in its architecture, but claims to be much more efficient than both Pung and Pung+}
\section{Related Work}
\label{sec:related}
2PPS provides a publish/subscribe protocol that achieves provable publisher and subscriber unobservability against a global active adversary who may also corrupt arbitrary clients and all but one server.
Privacy protection is achieved via a combination of DPF-based secret sharing for publishing and PIR for subscribing to topics.
In this section, we provide an overview of existing anonymous communication protocols and how they compare to 2PPS.

\paragraph*{Mixnets-based Protocols.}
Mix networks (mix nets) \cite{RefMix1, RefMix2, RefMix3} ensure the anonymity of users by obfuscating the source of a message.
They work by collecting the messages from many users and shuffle them by a set of servers called mixes before sending them out to recipients.
Therefore, they make it difficult for the global adversary to correlate input and output messages and protect against traffic analysis attacks.
However, malicious mixes can launch several attacks to deanonymize users; for instance, they can drop, modify, or duplicate the input messages before sending them out \cite{RefMixAtt1, RefMixAtt2, RefMixAtt3}.
Verifiable shuffles techniques \cite{RefMixVS1, RefMixVS2, RefMixVS3} have been proposed to protect against tampering messages by malicious servers, but these techniques introduce high computation overhead.
Mixnets-based protocols like McMix \cite{RefMcmix}, Atom \cite{RefAtom} and XRD \cite{RefXRD} induce high latency to support large numbers of users. While protocols as Vuvuzela \cite{RefVuvuzela}, Stadium \cite{RefStadium}, Alpenhorn \cite{RefAlpenhorn}, and Karaoke \cite{RefKaraoke} achieve better performance than 2PPS, but they provide weaker differential privacy guarantees compared to the cryptographic guarantees as 2PPS does.
In practice, this means that an adversary learns more information the longer he observes the mixnet, which is not the case for 2PPS.

\paragraph*{DCnets-based Protocols.}
Chaum's Dining Cryptographers network (DCnet) \cite{RefDCnet1, RefDCnet2} is an information-theoretic anonymous broadcast method.
To increase the scalability and practicality of DCnets, many protocols like Dissent \cite{RefDissent} and Verdict \cite{RefVerdict} adopted the client-server paradigm, where $n$ clients form the anonymity set, but only a small set of $N$ servers implement the functionality.
This adoption reduces the overall communication complexity from $O(n^2)$ in the traditional DCnet model (where there is a full graph between clients) to $O(N\cdot n)$ in the client-server DCnet model.
The sender anonymity guarantees that DCnet protocols provide are the same as 2PPS does.
However, the size of the sent and the received message in 2PPS is significantly smaller than in DCnet protocols.
That is because 2PPS uses DPF to compress the write requests and IT-PIR to only retrieve the blocks that the client is interested in, instead of getting all the messages through broadcasting.
Also, the used primitives in our protocol allow it to support a much larger number of clients and provide faster communication time than DCnet protocols do \cite{RefRiposte, RefExpress}.  

\paragraph*{PIR-based Protocols.}
Many protocols rely on PIR methods to enable anonymous communication.
There are two classes of these protocols that are: 1) information theoretic-PIR-based protocols (multiserver) such as Express \cite{RefExpress}, Riposte \cite{RefRiposte}, Blinder \cite{RefBlinder}, and Talek \cite{RefTalek}; and 2) computational-PIR-based protocols (one server) such as Pung \cite{RefPung}.
Express uses DPF to allow a user to send a message anonymously to the mailbox of another user.
However, this protocol does not protect the anonymity of the mailbox's owner (i.e., the receiver's anonymity).
Riposte and Blinder also use DPF but broadcast all the published messages.
As we have shown in our evaluation (\Cref{sec:eval}), broadcasting results in much higher network overhead than 2PPS's PIR, making it not suitable for bandwidth-restricted scenarios.
DPF-based protocols in general have a comparatively low computational overhead and can scale to support millions of users.
Talek provides a private publish-subscribe protocol that allows communication between small groups of trusted users.
In contrast, 2PPS's overhead does not depend on the number of subscribers a topic has and users neither have to trust publishers nor other subscribers to protect their privacy.
Pung operates in a single-server setting and provides strong anonymity guarantees since it can hide user's interests even if the server is malicious.
However, it introduces more overhead than 2PPS and requires users to trust their communication partners.

\section{Discussion}
\label{sec:disc}
In this section, we present some lessons learned from designing a public-message protocol based on secret sharing and private information retrieval.
\paragraph*{Availability}
As discussed in \Cref{sec:arch}, 2PPS owes its strong provable privacy protection to the combination of secret sharing and private information retrieval.
Both of these techniques distribute trust by relying on multiple servers.
This distribution of trust is great from a privacy perspective:
As long as one server remains honest, user privacy is preserved.
Users with high privacy requirements can even deploy their own servers to increase the chance of an honest one existing.

While very advantageous when it comes to privacy, secret sharing and PIR are very vulnerable as far as availability is concerned:
If a \emph{single} malicious server refuses to provide its write-database state after clients sent their messages, the protocol execution cannot continue.
To reveal the messages, \emph{all} shares are required by design.
The same problem arises on the reading site:
If a malicious server refuses to send his PIR response, the clients will not be able to receive the message from their subscribed topics.
Thus, as related literature~\cite{RefTalek, RefRiposte}, we assume that malicious servers do not target availability.
\begin{remark}[Backup Servers]
    An obvious mitigation against availability attacks from servers would be to introduce a ``backup'' server for each server.
    The backup server would receive the same information as its corresponding main server.
    In case the main server refuses to submit his shares, the backup server can step in.
    This avoids disruption as long as not both one main server and its backup refuse to participate.
    However, this comes at the cost of additional trust.
    Instead of only requiring one server to be honest, two honest servers are required;
    One honest main server and one honest backup server.
    Further, communication overhead for publishing messages also increases twofold, since the number of servers is doubled.
\end{remark}
Malicious clients may also target availability.
While the auditing protocol prevents clients from submitting malformed requests that would corrupt the write database, there are other avenues of attack which are inherent to 2PPS' open nature:
An adversary could create a large number of topics, increasing the required size of the write- and read databases and therefore also the network overhead of the whole system.
Further, the adversary could also spawn a large number of clients who all submit messages to a single topic, increasing the amount of cover needed for all other topics.
In scenarios where availability is of greater concern than the open nature of the systems, mechanisms can be put in place that increase the effort of adding users and topics to the system.

\paragraph*{Mitigating Intersection Attacks}
Due to the public nature of messages and topics, 2PPS is particularly vulnerable to a specific kind of intersection attack:
Assume that the adversary $\mathcal{A}$ wants to find out the interests of client Alice, who only publishes to a single topic.
Every time Alice is participating in a communication round, $\mathcal{A}$ records which topics are active (i.e., have messages sent to them).
Over multiple rounds, $\mathcal{A}$ \emph{intersects} the list of active topics until only a single topic remains, which is unambiguously linked to Alice.

Intersection attacks are inherently possible in any protocol that allows users to choose when they want to participate (see \Cref{sec:app:intersection} for proof).
While we assume constant participation for our security analysis, we also present possible mitigation techniques against intersection attacks for situations where constant participation is not obtainable:
\begin{itemize}
    \item \textbf{Cover Traffic.}
        A simple mitigation that is already in use with 2PPS is cover traffic.
        If Alice uses cover traffic, then $\mathcal{A}$ cannot distinguish rounds where Alice is sending ``real'' messages from rounds where she is participating idly, increasing the number of rounds $\mathcal{A}$ needs to observe.
    \item \textbf{Delayed Publishing.}
       Alice can also require that her message is not published in the round where she sent it but later (when Alice might already be offline).
        To do so, Alice can include some random delay $d$ in the message-topic tuple $t\mid m$ and \emph{encrypt} it with the public key of one of the servers.
        After combining their $D_w$ states, the corresponding server will reveal the ciphertext and delay publishing it. This solution is based on the idea of distributing knowledge which means each server knows the real publishing time of only a subset of all messages. Thus, $\mathcal{A}$ who controls one of the servers cannot accurately link every message to the set of the potential senders.  
\end{itemize} 

\paragraph*{Collisions}
In 2PPS, the client chooses a random row in the database to write her message into. 
Therefore, it is possible to have collisions, i.e., two or more clients writing their messages in the same row which corrupts both messages. 
To minimize the probability of collisions during normal operation, 2PPS uses a write database that is much larger than the number of participating clients.
However, this solution cannot solve the problem completely. 
If a client does not find her message among the published message in a given round, she has to assume that a collision occurred and may try to send her message again in a later round. 
Blinder~\cite{RefBlinder} employs another approach to deal with collisions:
If a client wants to publish message $m$, she computes one set of secret shares with $m$ at a random index and another set of shares with the square of $m$ at the same index.
The servers store the received shares in separate databases.
If a single collision occurred for a given index, the combined shares of both databases form a solvable system of quadratic equations, enabling the servers to reconstruct both original messages.
While this reduces the required storage space on the server side, it requires more computational overhead for clients and servers.
\section{Conclusion \& Future Work}
\label{sec:conc}
2PPS provides an anonymous publish/subscribe protocol with strong provable privacy guarantees for both publishers of messages and subscribers.
This is achieved by combining secret sharing based on distributed point functions for publishing with private information retrieval for accessing subscribed topics.
Compared to previous work, additional protection mechanisms are introduced to the secret sharing:
A combination of timestamps and encryption allows 2PPS to provide publisher privacy not only against a malicious server but also against stronger adversaries that can observe and interfere with traffic on all network links.
Our experimental evaluation shows that 2PPS can support a large number of users with latency suitable for applications such as microblogging and newsfeeds.
\begin{comment}
While 2PPS reaches its stated goals, we still see some avenues for future improvement. 
First, we aim to improve building blocks to lower latency so 2PPS can support use cases such as real-time video streaming. 
Second,  we will extend the functionalities of 2PPS to allow users to retrieve messages anonymously from previous rounds and subscribe to multiple topics using one subscription request. 
\end{comment}
While 2PPS reaches its stated goals, we see the following avenues for future improvement:
\begin{itemize}
    \item Improvements to the auditing protocol to increase the number of servers without introducing high computation and communication costs.
    \item Enabling users to subscribe to multiple topics using one subscription request. If servers would be able to handle all the subscriptions of one client at once, more efficient packing of messages may be possible, reducing the amount of cover needed.
    \item Allowing the anonymous retrieval of messages from previous rounds.
    \item Lowering latency to enable use cases such as live streaming.
\end{itemize}

\subsection*{Acknowledgments}
We thank Martin Byrenheid, Clemens Deusser, and Ephraim Zimmer for their valuable feedback and discussion.

\appendix

\section{Appendix}
\subsection{Proving Sender Unobservability}
\label{sec:app:so}
\begin{lemma}[Message Unlinkability]
    \label{lem:ml}
    $\mathcal{A}$ cannot identify which sender sent a given message.
\end{lemma}
\begin{IEEEproof}
    We proof \Cref{lem:ml} by showing that $\mathcal{A}$ cannot use any of his abilities to do so.
    \begin{itemize}
        \item \textbf{Passive Observation.}
            Every client sends one request per round to each server.
            All requests are encrypted under the receiving servers public key.
            The servers collect the incoming requests and reveal all messages from the current round at once.
            Thus, $\mathcal{A}$ cannot link messages to senders by passively observing requests between clients and servers.
        \item \textbf{Server Corruption.}
            According to our assumptions, $\mathcal{A}$ is able to corrupt all but one 2PPS servers.
            $\mathcal{A}$ can only match a client to the request he sends prior to adding it to the $db_w$ state.
            However, at this point, $\mathcal{A}$ can learn at most $N-1$ of the clients $N$ DPF-shares.
            Related literature has formal proves that combining $N-1$ shares does not reveal any information about the enclosed message \cite{RefDPF}.
        \item \textbf{Replay.}
            $\mathcal{A}$ could either replay request during the same or a later round.
            If $\mathcal{A}$ replays a request during the same round as the original request was sent, the honest server will detect identical requests arriving and discard all but one.
            If $\mathcal{A}$ replays a request during some later round, the honest server will notice that the included timestamp is not valid and discard the request.
            $\mathcal{A}$ is not able to update the timestamp of the replayed request, since it is protected by an encryption layer prior to the honest server.
        \item \textbf{Modification.}
            To be able to link a request to the client who sent it, $\mathcal{A}$ has to modify the request prior to the honest server.
            Since the contained DPF-share is encrypted, any modification at this point will lead to unpredictable changes of the share.
            Such a share will be rejected by the honest server's auditing protocol with overwhelming probability.
        \item \textbf{Dropping.}
            We assume that the honest server can detect a dropped request and refuse further protocol participation.
    \end{itemize}
\end{IEEEproof}

\begin{theorem}[Sender Unobservability]
    2PPS achieves Sender Unobservability.
\end{theorem}
\begin{IEEEproof}
    We define a series of hybrid games:
    \begin{itemize}
        \item $H_0$: The original $S\overline{O}$ game
        \item $H_1$: $H_0$, but clients can only publish 0-messages to a random topic
        \item $H_2$: $H_1$, but clients can only publish cover messages
        \item $H_3$: Identical Scenarios
    \end{itemize}
    In the following, we show that any adversary that can win $H_i$ non-negligible advantage can also win $H_{i+1}$ with non-negligible advantage.
    \begin{itemize}
        \item $H_0 \approx H_1$:
            We assume that $\mathcal{A}$ can win $H_0$.
            \Cref{lem:ml} states that $\mathcal{A}$ does so without being able to link any revealed message to a sender.
        \item $H_1 \approx H_2$:
            The only difference between a cover message and a ``real'' message is it's content:
            While a real message is sent to a chosen topic and can contain an arbitrary plaintext, a cover message is sent to a random topic and contains only 0s.
            We note that $H_1$ already requires real messages to have identical content to cover messages.
            Thus, all messages in $H_1$ are indistinguishable from cover messages to $\mathcal{A}$.
            If $\mathcal{A}$ can win $H_1$, he can therefore also win $H_2$.
        \item $H_2 \approx H_3$:
            As mentioned previously, we assume that all clients participate in every round.
            Further, every client sends exactly one message in each round.
            In $H_2$, all messages are sent to a random topic and contain only 0.
            Thus, $\mathcal{A}$ already has no influence on protocol activity in $H_2$ and the scenarios therefore appear identical to him.
    \end{itemize}
    We have shown that any adversary who can win $H_0$ can also win $H_3$ with a non-negligible advantage.
    Since $\mathcal{A}$ is only allowed to submit identical scenarios in $H_3$, there cannot be any such $\mathcal{A}$. 
    Therefore, no $\mathcal{A}$ can win $H_0$, which is equivalent to the $S\overline{O}$ game.
\end{IEEEproof}
 
\subsection{Proving Receiver Unobservability}
\label{sec:app:ro}
\begin{theorem}[Receiver Unobservability]
    2PPS achieves Receiver Unobservability.
\end{theorem}
\begin{IEEEproof}
    With Receiver Unobservability, $\mathcal{A}$ may not learn any information about receiver activity.
    $\mathcal{A}$ can either gain information about receivers by observing subscription registrations or communication.
    \begin{itemize}
        \item \textbf{Subscription Registration.}
            Each client sends his subscription registrations at a fixed rate to each server.
            Prior to the receiving server, the registration is always encrypted under the public key of this server.
            An active adversary may be able to corrupt a clients subscription registration.
            However, this does not lead to differing behavior based on the topic the client wants to subscribe to, since receivers show do not react to the messages they receive.
            
            We assume that $\mathcal{A}$ can corrupt $N-1$ servers.
            This does not help him in determining the topic a client is subscribing to, since any combination of $N-1$ requests per definition appears random to $\mathcal{A}$.
            Only the combination of all $N$ requests reveals the topic.
        \item \textbf{Communication.}
            Since the subscription request appears random to every server, no combination of $N-1$ servers can determine which topic the client is subscribed to by computing the response $res_i$.
            Although the primary server $P$ has access to all individual responses, it cannot reveal the messages, since they are obfuscated by the shared secret $s_j$ between the client and the honest server.
    \end{itemize}
\end{IEEEproof}
\subsection{Intersection Attacks}
\label{sec:app:intersection}
\begin{definition}[Delivery-Guarantee]
    A protocol provides Delivery-Guarantee, if all messages that are sent in a given round are also published in the same round.
\end{definition}
\begin{definition}[Sending-Nonobligation]
    A protocol provides Sending-Nonobligation, if it does not enforce when users send messages.
\end{definition}
Common approaches against intersection attacks~\cite{RefBuddies} do \emph{not} provide Sending-Nonobligation.
\begin{theorem}
    \label{thm:intersection}
    A protocol that provides both Delivery-Guarantee and Sending-Nonobligation cannot provide sender-messages unlinkability against an adversary who learns which messages are sent if the protocol guarantees delivery of all messages sent in a given round.
\end{theorem}
\begin{IEEEproof}
    Without Sending-Nonobligation, clients may only participate in a subset of all rounds.
    $\mathcal{A}$ observes a messages $m$ being published in round $r$
    Due to Delivery-Guarantee, a client that has not participated in round $r$ cannot have been the sender of $m$.

    If $\mathcal{A}$ can determine that another message $m^\prime$ was sent by the same sender, he can further narrow down the set of possible senders of $m$ and $m^\prime$ via an \emph{intersection} attack.
\end{IEEEproof}

\end{document}